\documentclass[twocolumn,pra,showpacs,nofootinbib,groupedaddress]{revtex4}
\usepackage{amssymb}
\usepackage{amsmath}
\usepackage{amsfonts}

\usepackage{dcolumn,graphicx}

\begin{document}

\title{Micromagic clock: microwave clock based on atoms in an engineered optical lattice}

\author{K. Beloy}
\affiliation {
Department of Physics, University of Nevada, Reno, Nevada 89557}
\author{A. Derevianko}
\affiliation {
Department of Physics, University of Nevada, Reno, Nevada 89557}
\author{V. A. Dzuba}
\affiliation {
School of Physics, University of New South Wales, Sydney,
2052, Australia}
\author{V. V. Flambaum}
\affiliation {
School of Physics, University of New South Wales, Sydney,
2052, Australia}
%\affiliation {
%Perimeter Institute for Theoretical Physics
%31 Caroline Street North,
%Waterloo, Ontario  N2L 2Y5, Canada}

\date{\today}

\begin{abstract}
We propose a new class of atomic microwave clocks based on the
hyperfine transitions in the ground state of aluminum or gallium
atoms trapped in optical lattices. For these elements {\em magic}
wavelengths exist at which both levels of the hyperfine doublet are
shifted at the same rate by the lattice laser field, cancelling its
effect on the clock transition. Our analysis of various systematic
effects shows that, while offering an improved control over
systematic errors, the accuracy of the proposed microwave clock is
competitive to that of the state-of-the-art primary frequency
standard.
\end{abstract}

% 06.30.Ft Time and frequency
% 37.10.Jk Atoms in optical lattices
% 31.15.A- <i>Ab initio</i> calculations
\maketitle

The present definition of the unit of time, the second, is based on
the frequency of the microwave transition between two hyperfine
levels of Cs atom. The Cs atomic clocks date back more than a half
of a century. The accuracy of the standard has been substantially
improved over the years culminating in a fountain clock apparatus
operated around the world~\cite{BizLauAbg05etal,HeaJefDon05etal}.
Recently,  it has been realized that the accuracy and stability of
atomic clocks can be substantially improved by trapping atoms in a
standing wave of a laser light (optical lattices) operated at a
certain ``magic'' wavelength~\cite{KatTakPal03,YeKimKat08}. The
laser wavelength is tuned so that the differential light
perturbations of the two clock levels vanishes exactly. In other
words, while remaining confined (this eliminates the Doppler and
recoil shifts), the atoms behave spectroscopically as if they were
in a vacuum. Millions of atoms can be trapped and interrogated
simultaneously, vastly improving stability of the clock. Such setup
was experimentally
realized~\cite{TakHonHig05,LeTBaiFou06,LudZelCam08etal} for optical
frequency clock transitions in divalent atoms yielding accuracies
competitive to the fountain clocks~\cite{LudZelCam08etal}. However,
because these lattice clocks operate at an optical frequency, to
relate to the definition of the second, they require state of the
art frequency combs to bridge the optical frequency to the microwave
counters.

Here we extend the fruitful ideas of the optical lattice clocks to
microwave frequencies. We propose a new class of atomic microwave
clocks based on the hyperfine transitions in the ground state of
aluminum or gallium atoms trapped in optical lattices. We determine
the  magic wavelengths and analyze various systematic effects. We
find that the accuracy of the proposed clockwork is at least
competitive to that of the most-precise clocks. Moreover, compared
to a large chamber of the fountain clock, the atoms are confined to
a tiny volume offering improved control over systematic errors. A
relative compactness of the clockwork could benefit spacecraft
applications such as navigation systems and precision tests of
fundamental theories.

The microwave clockwork involves two atomic levels of the same
hyperfine manifold. The transition frequency is monitored and
ultimately translated into a time measurement. We envision the
following experimental setup: the atoms are trapped in a 1D optical
lattice formed by counter-propagating laser beams of linear
polarization and frequency $\omega_{L}$. The quantizing magnetic
field $\mathbf B$ is directed either along the direction of the
laser propagation $\hat k$ or along the polarization vector $\hat
\varepsilon$. The clock states are commonly labeled as $\left\vert
n\left( IJ\right) FM_{F}\right\rangle $, where $I$ is the nuclear
spin, $J$ is the electronic angular momentum, and$\ F$ is the total
angular momentum, $\mathbf{F}=\mathbf{J}+\mathbf{I}$, with $M_{F}$
being its projection on the quantization axis and $n$ encompassing
the remaining quantum numbers.  To minimize a sensitivity to stray
magnetic fields and residual circular polarization of the laser
light, we choose to work with the $M_{F}=0$ components.

Under the influence of the laser each clock level is perturbed. The
relevant energy shifts are parameterized in terms of the dynamic
scalar,$~\alpha_{a}^{S}\left(  \omega_{L}\right)  $, and tensor,
$\alpha_{a}^{T}\left(  \omega_{L}\right) $, polarizabilities.
The corresponding expression reads%
\begin{eqnarray}
\lefteqn{\delta E_{n\left(  IJ\right)  FM_{F}}^{\mathrm{Stark}}\left(  \omega
_{L}\right)  =-\left(  \frac{1}{2}\mathcal{E}_{0}\right)  ^{2}\left[
\alpha_{n\left(  IJ\right)  F}^{S}\left(  \omega_{L}\right)  \right.} \nonumber \\
&& \left.
 + \xi(\theta) \, \alpha_{n\left(
IJ\right)  F}^{T}\left(  \omega_{L}\right)  \frac{3M_{F}^{2}-F\left(
F+1\right)  }{F\left(  2F-1\right)  }\right]  ,
\end{eqnarray}
where $\xi(\theta)=(3 \cos^2\theta-1)/2$, $\theta$ being the angle
between $\mathbf B$ and $\hat \varepsilon$. In particular, $\xi=1$
for $\mathbf B \parallel \hat \varepsilon$ and $\xi=-1/2$ for
$\mathbf B\parallel \hat k$ geometries. $\mathcal{E}_{0}$ is the
amplitude of the laser field. While arriving at these expressions we
required that the Zeeman splittings in the B-field are much larger
than the off-diagonal matrix of the optical Hamiltonian, a condition
which can be easily attained experimentally.

The clock frequency is modified by the difference
\[
\delta\nu^{\mathrm{Stark}}\left(  \omega_{L}\right)  =\frac{1}{h}\left(
\delta E_{n\left(  IJ\right)  F^{\prime\prime}M_{F}^{\prime\prime}%
}^{\mathrm{Stark}}\left(  \omega_{L}\right)  -\delta E_{n\left(  IJ\right)
F^{\prime}M_{F}^{\prime}}^{\mathrm{Stark}}\left(  \omega_{L}\right)  \right)
\]
We require that at a certain, \textquotedblleft magic\textquotedblright, laser
frequency $\omega_{L}^{\ast}$, this laser-induced differential shift
vanishes: $\delta\nu\left(  \omega_{L}^{\ast}\right) =0$.
%The frequency depends on the mutual orientation of quantizing B field and the %k-
%polarization vector of the laser.

\begin{figure}[h]
\begin{center}
\includegraphics*[scale=0.4]{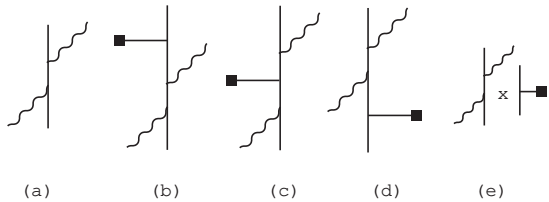}
\end{center}
\caption{Contributions to the dynamic polarizability $\alpha(\omega)$. Interactions
with the laser photons are shown with wavy lines and hyperfine interaction
with the capped solid line. (a) Second-order $\alpha^{(2)}(\omega)$.
Contributions (b)--(e) are the
third-order contributions to polarizabilities (b) top, (c) center, (d) bottom,
and (e) normalization diagrams.}
\label{Fig:diagrams}
\end{figure}

The ``magic'' cancellation mechanism depends on the frequency
dependence of underlying polarizabilities. We use perturbation
theory and expand the polarizabilities in terms of the powers of the
hyperfine interaction $\alpha_{a}\left(  \omega\right)  =\alpha
_{a}^{\left(  2\right)  }\left(  \omega\right)  +\alpha_{a}^{\left(
3\right) }\left(  \omega\right)  +\ldots$ The leading term,
$\alpha_{a}^{\left( 2\right)  }\left(  \omega\right)  $, contains
the interaction with two photons and $\alpha_{a}^{\left(  3\right)
}\left(  \omega\right)  $ in addition involves the hyperfine
coupling of the electrons with the nuclear spin. The relevant
diagrams are shown in Fig.~\ref{Fig:diagrams}. It is important to
realize that the scalar component of $\alpha_{a}^{\left(  2\right)
}\left(  \omega\right)  $ does not depend on $F$. Also for $J=1/2$
levels, the tensor component of $\alpha _{a}^{\left(  2\right)
}\left(  \omega\right)  $ vanishes due to selection rules, as the
underlying rotational symmetry is that of the rank 2 tensor. We
conclude that for a $J=1/2$ level in a laser light of linear
polarization, the dominant Stark shift vanishes. Consequently, below
we restrict consideration to the $J=1/2$ levels.

Since in the leading order both clock levels are shifted
identically, we proceed to computing the third-order diagrams of
Fig.~\ref{Fig:diagrams}. Each diagram involves two couplings to the
lattice laser field and one hyperfine interaction (HFI). The
labeling of the diagrams (top, center, and bottom) reflects the
position of the HFI with respect to the two laser interactions. The
formalism and the computational scheme is similar to those of
Refs.~\cite{BelSafDer06,AngDzuFla06PRL}. We carry out the
conventional angular reduction and extract the scalar and tensor
contributions to each diagram. We find that the third-order shift
may be parameterized as
\begin{eqnarray}
\delta\nu^{\mathrm{Stark}}\left(  \omega_{L}\right)  &=&\left(
\frac{1}{2}\mathcal{E}_{0}\right)  ^{2}\left(  A\left(  F^{\prime}%
,F^{\prime\prime}\right)  \left[  \alpha_{n\left(  IJ\right)  F^{\prime}%
}^{\left(  3\right)  }\left(  \omega_{L}\right)  \right]  ^{\mathrm{Scalar}%
}
\right.
\nonumber \\
&+& \left. B\left(  F^{\prime},F^{\prime\prime}\right)  \left[  \alpha_{n\left(
IJ\right)  F^{\prime}}^{\left(  3\right)  }\left(  \omega_{L}\right)  \right]
^{\mathrm{Tensor}}\right)  , \label{Eq:ClockShiftSandT}%
\end{eqnarray}
where coefficients $A\left(  F^{\prime},F^{\prime\prime}\right)  $ and
$B\left(  F^{\prime},F^{\prime\prime}\right)$ depend on the $F$-numbers of
the clock states and on the orientation ($\parallel$ or $\perp$) of the
quantizing B-field with respect to the %propagation of the laser.
polarization vector of the laser light. The relation
(\ref{Eq:ClockShiftSandT}) arises due to the fact that the
respective scalar and tensor parts of the dynamic polarizability
%$\alpha_{a}^{\left( 3\right) }\left( \omega\right) $
vary proportionally for the two clock states. Clearly the scalar and
tensor contributions to the differential shift must cancel each
other at the magic wavelength.

We start with discussing the results for the metrologically
important $^{133}$Cs atom. A lattice Cs microwave clock was
discussed in Ref.~\cite{ZhoCheChe05}. Here the clock transition is
between the $F=4$ and $F=3$ hyperfine components of the electronic
ground state $6s_{1/2}$. Since $J=1/2$, for linear polarization the
second-order shift of the clock frequency vanishes and we need to
proceed to the third-order diagrams, Fig.~\ref{Fig:diagrams}
(b)-(e). We carried out relativistic many-body calculations of these
diagrams and found that there is no magic wavelength for the Cs
clock. This is in contrast to findings of Ref.~\cite{ZhoCheChe05},
where a multitude of magic wavelengths was identified.
Qualitatively, for Cs, the tensor contribution to the clock shift is
much smaller than the scalar contribution and this leads to
unfavorable conditions for reaching the cancellation of the scalar
and tensor shifts in Eq.~(\ref{Eq:ClockShiftSandT}).

We conclude that to cancel the third-order light shift we need to
find atoms where the scalar and tensor shifts are comparable. This
happens for  atoms having the valence electrons in the $p_{1/2}$
state. For non-zero nuclear spin, the $p_{1/2}$ state has two
hyperfine components that may serve as the clock states. Moreover,
since the electronic angular momentum $J=1/2$ for the linear
polarization the leading second-order shift of the clock frequency
vanishes. This is similar to the Cs $s_{1/2}$ case. The advantage of
the $p_{1/2}$ state comes from the fact it is part of a
fine-structure manifold: there is a nearby $p_{3/2}$ state separated
by a relatively small energy interval determined by the relativistic
corrections to the atomic structure. The hyperfine interaction
between the states of the same manifold is amplified due to small
energy denominators entering top and bottom diagrams of
Fig.~\ref{Fig:diagrams}. The amplification occurs only for the
tensor contribution. For the scalar contribution the intermediate
state must be of the of $p_{1/2}$ symmetry,
%while for the tensor contribution the additional, strongly coupled,
%$p_{3/2}$ channel becomes open.
whereas for the tensor contribution the intermediate state must be
of the (strongly coupled) $p_{3/2}$ symmetry.

We illustrate this qualitative discussion with numerical examples
for the group III atoms. We start with aluminum ($Z=13$). The clock
transition is between the hyperfine structure levels $F=3$ and $F=2$
in the ground $3p_{1/2}$ state of $^{27}$Al isotope ($I=5/2$). The
clock frequency has been measured to be 1.50614(5)
GHz~\cite{LewWes53}, placing it in the microwave region. The
$\mu$Magic clock setup requires ultracold atoms. Cooling Al has
already been demonstrated \cite{McGGilLee95} with the goal of atomic
nanofabrication. The laser cooling was carried out on the closed
$3p_{3/2}-3d_{5/2}$ transition with the recoil limit of 7.5 $\mu$K.
Once trapped, the atoms can be readily transferred from the
metastable $3p_{3/2}$ cooling state to the ground (clock) state.
Lattice-trapped Al was also considered for quantum information
processing~\cite{RavDerBer06}.
%
%$\mathbf B \parallel \hat \varepsilon$
% $\mathbf B\parallel \hat k$

Using relativistic many-body theory we computed the polarizabilities
for the two experimental geometries ($\mathbf B\parallel \hat k$ or
$\mathbf B \parallel \hat \varepsilon$) and found three magic
frequencies. For  the $\mathbf B\parallel \hat k$ configuration the
magic wavelengths and the second-order polarizabilities are
\begin{align*}
\lambda_{L}^{\ast} &  =390\text{ nm, }\alpha^{S}\left(  \omega_{L}^{\ast
}\right)  =-211\text{ a.u.,}\\
\lambda_{L}^{\ast} &  =338\text{ nm, }\alpha^{S}\left(  \omega_{L}^{\ast
}\right)  =+142\text{ a.u.,}%
\end{align*}
and for the $\mathbf B \parallel \hat \varepsilon$ geometry
\[
\lambda_{L}^{\ast}=400\text{ nm, }\alpha^{S}\left(  \omega_{L}^{\ast}\right)
=+401\text{ a.u.}%
\]
The $3p_{1/2}-4s_{1/2}$ transition is 394.5 nm, indicating a
blue-detuning for the first two ($\mathbf B\parallel \hat k$) magic
wavelengths and a red-detuning for the last one ($\mathbf B
\parallel \hat \varepsilon$).

\begin{figure}[h]
\begin{center}
\includegraphics*[scale=0.4]{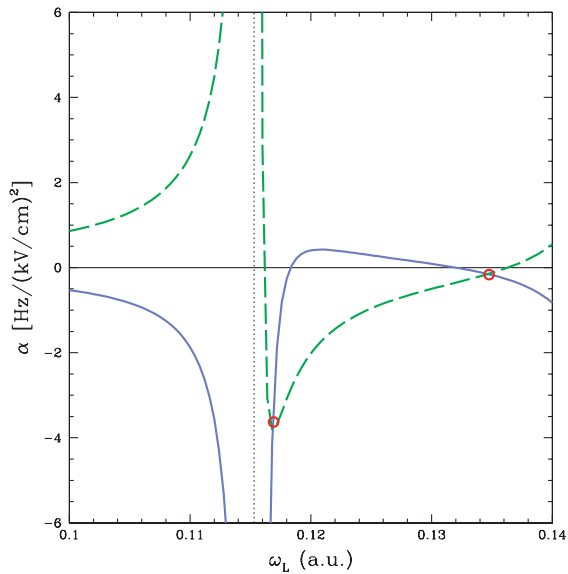}
\end{center}
\caption{(Color online) Third-order shift of the clock levels
$|F=3,M_F=0\rangle$ (dashed line) and $|F=2,M_F=0\rangle$ (solid
line) for Al $\mu$Magic clock  in the $\mathbf B \parallel \hat k$
geometry as a function of the lattice laser frequency. The shifts
are identical at the magic frequencies (red circles) above the
$3p_{1/2}-4s_{1/2}$ resonance (vertical dotted line).
%solid line: F=2, long dashed line: F=3, dotted line: resonance position (3p
%1/2 - 4s, w=0.1153 a.u.).
}
\label{Fig:AlClockShifts}
\end{figure}

\begin{figure}[h]
\begin{center}
\includegraphics*[scale=0.4]{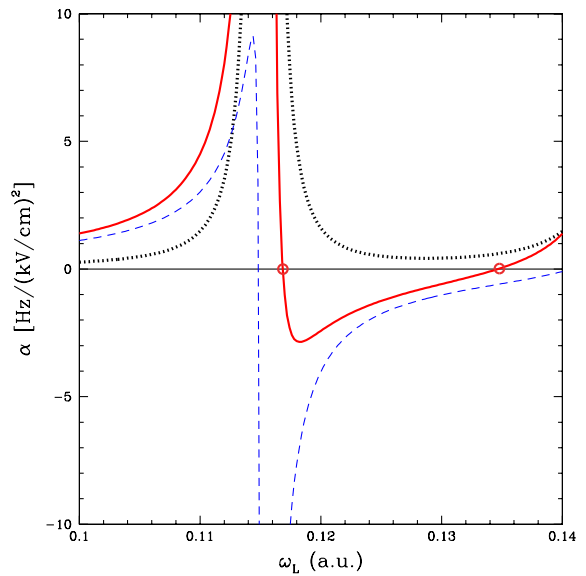}
\end{center}
\caption{(Color online) Differential polarizability for Al
$\mu$Magic clock in the $\mathbf B \parallel \hat k$  geometry as a
function of the lattice laser frequency. Dotted line: contribution
from scalar term; dashed line: contribution from the tensor term;
solid line: total polarizability. Total clock shift vanishes at two
values of the laser frequency. } \label{Fig:AlCancellation}
\end{figure}

The third-order Stark shifts of the clock levels as a function of
$\omega_L$ are shown in Fig.~\ref{Fig:AlClockShifts}. At the magic
wavelength the Stark shifts are identical and the clock transition
is unperturbed. The cancellations between scalar and tensor
contributions in Eq.~(\ref{Eq:ClockShiftSandT}) to the clock shift
is illustrated in Fig.~\ref{Fig:AlCancellation}.

The values of polarizability $\alpha^{S}\left(
\omega_{L}^{\ast}\right)  $ determine the depths of the optical
potentials. In general, a laser of intensity 10 kW/cm$^{2}$ would be
able to hold atoms of temperature 10 $\mu K$. The atoms are trapped
in the intensity minima of the standing wave for $\alpha^{S}\left(
\omega_{L}^{\ast}\right)  <0$ and in the maxima otherwise. Both
cases are realized depending on the geometry. For the blue-detuned
case, one could use hollow beams to confine atoms in the radial
direction. We also notice that a 3D lattice can be formed by three
sets of counter-propagating laser beams. In this case one pair
operates in the $\mathbf B \parallel \hat k$ geometry and with the
appropriate magic wavelength, while the orthogonal beams operate at
the $\mathbf B \parallel \hat \varepsilon$ magic wavelength.

Presently, the factor limiting the accuracy of the most precise
neutral-atom clocks is the black-body radiation (BBR), which arises
due to an interaction of a thermal bath of photons at ambient
temperature $T$ with the clock~\cite{PorDer06BBR,LudZelCam08etal}.
Due to the isotropic and low-frequency character of the BBR, the
relevant shift involves only the scalar part of the polarizability
evaluated in the static limit and given by the third-order diagrams
of Fig.~\ref{Fig:diagrams} (see formalism of
Refs.~\cite{BelSafDer06,AngDzuFla06PRL}). The fractional
contribution is conventionally parameterized as
\[
\frac{\delta\nu^{BBR}}{\nu_{0}}=\beta~\left(  \frac{T}{300}\right)  ^{4},
\]
where $T$ is expressed in Kelvins. We find that our computed
coefficient $\beta\left(  ^{27}\mathrm{Al}\right)
=-8.7\times10^{-16}$ is about 20 times smaller than the coefficient
for the Cs standard. In other words, the Al $\mu$Magic clock by an
order of magnitude is less susceptible to the BBR than the primary
standard. As $\beta$ may be determined from experiments with static
E-fields, we notice that ultimate accuracy of the clock is limited
by an inhomogeneity of the BBR temperature across the experimental
chamber. A typical inhomogeneity of $0.1$ K results in an estimate
of the fractional accuracy at $10^{-18}$. As the BBR shift depends
steeply on $T$, the entire experimental chamber could be cooled down
cryogenically reducing the uncertainty even further; here the small
volume of the chamber offers a distinct advantage over the fountain
clocks~\cite{LudZelCam08etal}.

While the choice of the $M_{F}=0$ substates eliminates the first-order Zeeman
shift, the sensitivity to B-fields comes through the second-order Zeeman
shift
%????? begin
which appears due to mixing of different hyperfine components by
$B$: The relative shift of the clock frequency is
$\delta\nu^{Zeeman}/\nu _{0}\approx 2/9~\left(  \mu_{B}B/ h\nu_0
\right)  ^{2}=1.9 ~\times~10^{-7}B^{2}$, where $B$ is expressed in
Gauss.
% For a typical 10 mG magnetic field the fractional shift is below $10^{-17}$.
This problem is similar to that in the fountain clocks (Cs, Rb,...),
where specific efforts to map the magnetic field over the flight zone are made.
However, since in the lattice the atoms are confined to a tiny volume, one could
control/shield the B-fields to a better degree than in the fountain clocks.

So far we assumed that the light is linearly polarized. In practice
there is always a small degree of circular polarization
$\mathcal{A}$ present experimentally. The residual circular
component leads to an undesired clock shift through axial (vector)
dynamic polarizability $\alpha^v$. This effect is equivalent to a
``pseudo-magnetic'' field along $\hat k$. For the $p_{1/2}$ clock
levels $\alpha^v$ arises already in the second order; we find it to
be in the order of 100 a.u. For the $M_F$=0 levels the relevant
clock shift is zero
%???? begin
in the first order in $\alpha^v$. However, the shift could appear in
the second order in $\mathcal{A} \alpha^v$ since the vector term
mixes different hyperfine components. For a typical circular
polarization $\mathcal{A}\sim 10^{-5}$ and a misalignment angle of
$10^{-2}$, the fractional frequency shift is just $10^{-21}$.

Atoms of Al are bosons and the collisional clock shifts may become of issue,
as in the fountain clocks~\cite{PerMarBiz02,SzyChaTie07}. The advantage of the lattice
clocks over the fountain clocks is that one could fill the lattice with no
more than one atom per site, strongly suppressing the interatomic interactions
and the associated shifts.

The scattering of the lattice laser photons leads to heating and reduces the
interrogation time. At 10 kW/cm$^{2}$ the heating rate is in the order of
10$^{-2}$ $\sec^{-1}$. The heating can be suppressed by using the blue-detuned
magic wavelength ($\lambda_{L}^{\ast}=390$ nm), when the atoms are trapped at
the intensity minima. This also reduces the effects of hyperpolarizability on
the clock shift and multi-photon ionization rates.

We have carried out a similar analysis for Ga atom (both $^{69}$Ga
and $^{71}$Ga isotopes; $I=3/2$), a member of the same group III of
the periodic table as Al. Cooling of this atom is pursued in atomic
nanofabrication~\cite{RehBocLee04,CamMarFaz06}. The clock transition is between
the hyperfine structure components $F=1$ and $F=2$ of the $4p_{1/2}$
ground state, and has been measured to be 2.6779875(10) GHz and
3.4026946(13) GHz for $^{69}$Ga and $^{71}$Ga,
respectively~\cite{LurPro56}. In contrast to Al, for this atom and
the $M_F=0$ sublevels, we have identified only a single magic
wavelength at 450 nm in the
 $\mathbf B \parallel \hat \varepsilon$
geometry. This is red-detuned from the $4p_{1/2}-5s_{1/2}$
transition frequency of 403.4 nm. We find $\alpha^{S}\left(
\omega_{L}^{\ast}\right) = 94$ a.u. and a very small BBR coefficient
$\beta\left(  ^{69,71}\mathrm{Ga}\right) =-6.63 \times 10^{-17}$. We
did not find the magic wavelengths for other group III atoms.

%??? begin
To summarize, we proposed Al and Ga microwave lattice clocks ($\mu$Magic clocks).
We calculated magic wavelengths for these clocks where the laser-induced
differential Stark shift vanishes. This is a result of the opposite sign contributions
of the scalar and tensor polarizabilities to the Stark shift. The tensor polarizability in
the $p_{1/2}$ electron state is enhanced due to the mixing of $p_{1/2}$
and $p_{3/2}$ states by the hyperfine interaction. A similar mechanism
for the magic wavelengths may work in microwave hyperfine transitions in
other atoms which have the fine structure multiplets in the ground state.
In atoms with the valence electron in the $s_{1/2}$ state (Cs, Rb, ...) the magic
wavelength is absent. Estimates of the uncertainties show that the $\mu$Magic clocks
 may  successfully compete with the state of the art
fountain clocks.
%??? end

We gratefully acknowledge motivating discussion with P. Rosenbusch
and invaluable advice of E.N. Fortson, S.A. Lee, H. Katori, J.
Weinstein, J. Ye, P. Hannaford, N. Yu, and C. Salomon. AD was
supported in part by the US Dept.~of State Fulbright fellowship to
Australia. This work was supported in part by the US National
Science Foundation, by the Australian Research Council and by the US
National Aeronautics and Space Administration under
Grant/Cooperative Agreement No. NNX07AT65A issued by the Nevada NASA
EPSCoR program.

%%%%%%%%%%%%%%%%%%%%%%%%%%%%%%%%%%%%%%%%%%%%%%%%%%%%%%%%%%%%%%%%
%\bibliography{all,add}

\begin{thebibliography}{19}
\expandafter\ifx\csname natexlab\endcsname\relax\def\natexlab#1{#1}\fi
\expandafter\ifx\csname bibnamefont\endcsname\relax
  \def\bibnamefont#1{#1}\fi
\expandafter\ifx\csname bibfnamefont\endcsname\relax
  \def\bibfnamefont#1{#1}\fi
\expandafter\ifx\csname citenamefont\endcsname\relax
  \def\citenamefont#1{#1}\fi
\expandafter\ifx\csname url\endcsname\relax
  \def\url#1{\texttt{#1}}\fi
\expandafter\ifx\csname urlprefix\endcsname\relax\def\urlprefix{URL }\fi
\providecommand{\bibinfo}[2]{#2}
\providecommand{\eprint}[2][]{\url{#2}}

\bibitem[{\citenamefont{{Bize} et~al.}(2005)\citenamefont{{Bize}, {Laurent},
  and {Abgrall {\em et al.}}}}]{BizLauAbg05etal}
\bibinfo{author}{\bibfnamefont{S.}~\bibnamefont{{Bize}}},
  \bibinfo{author}{\bibfnamefont{P.}~\bibnamefont{{Laurent}}},
  \bibnamefont{and} \bibinfo{author}{\bibfnamefont{M.}~\bibnamefont{{Abgrall
  {\em et al.}}}}, \bibinfo{journal}{J. Phys. B} \textbf{\bibinfo{volume}{38}},
  \bibinfo{pages}{S449} (\bibinfo{year}{2005}).

\bibitem[{\citenamefont{Heavner et~al.}(2005)\citenamefont{Heavner, Jefferts,
  Donley, and {\em et al.}}}]{HeaJefDon05etal}
\bibinfo{author}{\bibfnamefont{T.}~\bibnamefont{Heavner}},
  \bibinfo{author}{\bibfnamefont{S.}~\bibnamefont{Jefferts}},
  \bibinfo{author}{\bibfnamefont{E.}~\bibnamefont{Donley}}, \bibnamefont{and}
  \bibinfo{author}{\bibnamefont{{\em et al.}}}, \bibinfo{journal}{Metrologia}
  \textbf{\bibinfo{volume}{42}}, \bibinfo{pages}{411} (\bibinfo{year}{2005}).

\bibitem[{\citenamefont{Katori et~al.}(2003)\citenamefont{Katori, Takamoto,
  Pal'chikov, and Ovsiannikov}}]{KatTakPal03}
\bibinfo{author}{\bibfnamefont{H.}~\bibnamefont{Katori}},
  \bibinfo{author}{\bibfnamefont{M.}~\bibnamefont{Takamoto}},
  \bibinfo{author}{\bibfnamefont{V.~G.} \bibnamefont{Pal'chikov}},
  \bibnamefont{and} \bibinfo{author}{\bibfnamefont{V.~D.}
  \bibnamefont{Ovsiannikov}}, \bibinfo{journal}{Phys.\ Rev.\ Lett.}
  \textbf{\bibinfo{volume}{91}}, \bibinfo{pages}{173005}
  (\bibinfo{year}{2003}).

\bibitem[{\citenamefont{Ye et~al.}(2008)\citenamefont{Ye, Kimble, and
  Katori}}]{YeKimKat08}
\bibinfo{author}{\bibfnamefont{J.}~\bibnamefont{Ye}},
  \bibinfo{author}{\bibfnamefont{H.~J.} \bibnamefont{Kimble}},
  \bibnamefont{and} \bibinfo{author}{\bibfnamefont{H.}~\bibnamefont{Katori}},
  \bibinfo{journal}{Science} \textbf{\bibinfo{volume}{320}},
  \bibinfo{pages}{1734} (\bibinfo{year}{2008}).

\bibitem[{\citenamefont{Takamoto et~al.}(2005)\citenamefont{Takamoto, Hong,
  Higashi, and Katori}}]{TakHonHig05}
\bibinfo{author}{\bibfnamefont{M.}~\bibnamefont{Takamoto}},
  \bibinfo{author}{\bibfnamefont{F.~L.} \bibnamefont{Hong}},
  \bibinfo{author}{\bibfnamefont{R.}~\bibnamefont{Higashi}}, \bibnamefont{and}
  \bibinfo{author}{\bibfnamefont{H.}~\bibnamefont{Katori}},
  \bibinfo{journal}{Nature (London)} \textbf{\bibinfo{volume}{435}},
  \bibinfo{pages}{321} (\bibinfo{year}{2005}).

\bibitem[{\citenamefont{{Le Targat} et~al.}(2006)\citenamefont{{Le Targat},
  Baillard, Fouch\'{e}, Brusch, Tcherbakoff, Rovera, and
  Lemonde}}]{LeTBaiFou06}
\bibinfo{author}{\bibfnamefont{R.}~\bibnamefont{{Le Targat}}},
  \bibinfo{author}{\bibfnamefont{X.}~\bibnamefont{Baillard}},
  \bibinfo{author}{\bibfnamefont{M.}~\bibnamefont{Fouch\'{e}}},
  \bibinfo{author}{\bibfnamefont{A.}~\bibnamefont{Brusch}},
  \bibinfo{author}{\bibfnamefont{O.}~\bibnamefont{Tcherbakoff}},
  \bibinfo{author}{\bibfnamefont{G.~D.} \bibnamefont{Rovera}},
  \bibnamefont{and} \bibinfo{author}{\bibfnamefont{P.}~\bibnamefont{Lemonde}},
  \bibinfo{journal}{Phys. Rev. Lett.} \textbf{\bibinfo{volume}{97}},
  \bibinfo{eid}{130801} (\bibinfo{year}{2006}).

\bibitem[{\citenamefont{Ludlow et~al.}(2008)\citenamefont{Ludlow, Zelevinsky,
  and {Campbell {\em et al.}}}}]{LudZelCam08etal}
\bibinfo{author}{\bibfnamefont{A.~D.} \bibnamefont{Ludlow}},
  \bibinfo{author}{\bibfnamefont{T.}~\bibnamefont{Zelevinsky}},
  \bibnamefont{and} \bibinfo{author}{\bibfnamefont{G.~K.}
  \bibnamefont{{Campbell {\em et al.}}}}, \bibinfo{journal}{Science}
  \textbf{\bibinfo{volume}{319}}, \bibinfo{pages}{1805} (\bibinfo{year}{2008}).

\bibitem[{\citenamefont{Beloy et~al.}(2006)\citenamefont{Beloy, Safronova, and
  Derevianko}}]{BelSafDer06}
\bibinfo{author}{\bibfnamefont{K.}~\bibnamefont{Beloy}},
  \bibinfo{author}{\bibfnamefont{U.~I.} \bibnamefont{Safronova}},
  \bibnamefont{and}
  \bibinfo{author}{\bibfnamefont{A.}~\bibnamefont{Derevianko}},
  \bibinfo{journal}{Phys. Rev. Lett.} \textbf{\bibinfo{volume}{97}},
  \bibinfo{pages}{040801} (\bibinfo{year}{2006}).

\bibitem[{\citenamefont{Angstmann et~al.}(2006)\citenamefont{Angstmann, Dzuba,
  and Flambaum}}]{AngDzuFla06PRL}
\bibinfo{author}{\bibfnamefont{E.~J.} \bibnamefont{Angstmann}},
  \bibinfo{author}{\bibfnamefont{V.~A.} \bibnamefont{Dzuba}}, \bibnamefont{and}
  \bibinfo{author}{\bibfnamefont{V.~V.} \bibnamefont{Flambaum}},
  \bibinfo{journal}{Phys. Rev. Lett.} \textbf{\bibinfo{volume}{97}},
  \bibinfo{pages}{040802} (\bibinfo{year}{2006}).

\bibitem[{\citenamefont{Zhou et~al.}(2005)\citenamefont{Zhou, Chen, and
  Chen}}]{ZhoCheChe05}
\bibinfo{author}{\bibfnamefont{X.}~\bibnamefont{Zhou}},
  \bibinfo{author}{\bibfnamefont{X.}~\bibnamefont{Chen}}, \bibnamefont{and}
  \bibinfo{author}{\bibfnamefont{J.}~\bibnamefont{Chen}}
  (\bibinfo{year}{2005}), \eprint{arXiv:0512244}.

\bibitem[{\citenamefont{Lew and Wessel}(1953)}]{LewWes53}
\bibinfo{author}{\bibfnamefont{H.}~\bibnamefont{Lew}} \bibnamefont{and}
  \bibinfo{author}{\bibfnamefont{G.}~\bibnamefont{Wessel}},
  \bibinfo{journal}{Phys. Rev.} \textbf{\bibinfo{volume}{90}},
  \bibinfo{pages}{1} (\bibinfo{year}{1953}).

\bibitem[{\citenamefont{McGowan et~al.}(1995)\citenamefont{McGowan, Giltner,
  and Lee}}]{McGGilLee95}
\bibinfo{author}{\bibfnamefont{R.~W.} \bibnamefont{McGowan}},
  \bibinfo{author}{\bibfnamefont{D.~M.} \bibnamefont{Giltner}},
  \bibnamefont{and} \bibinfo{author}{\bibfnamefont{S.~A.} \bibnamefont{Lee}},
  \bibinfo{journal}{Opt. Lett.} \textbf{\bibinfo{volume}{20}},
  \bibinfo{pages}{2535} (\bibinfo{year}{1995}).

\bibitem[{\citenamefont{Ravaine et~al.}(2006)\citenamefont{Ravaine, Derevianko,
  and Berman}}]{RavDerBer06}
\bibinfo{author}{\bibfnamefont{B.}~\bibnamefont{Ravaine}},
  \bibinfo{author}{\bibfnamefont{A.}~\bibnamefont{Derevianko}},
  \bibnamefont{and} \bibinfo{author}{\bibfnamefont{P.~R.}
  \bibnamefont{Berman}}, \bibinfo{journal}{Phys. Rev. A}
  \textbf{\bibinfo{volume}{74}}, \bibinfo{eid}{022330} (\bibinfo{year}{2006}).

\bibitem[{\citenamefont{Porsev and Derevianko}(2006)}]{PorDer06BBR}
\bibinfo{author}{\bibfnamefont{S.~G.} \bibnamefont{Porsev}} \bibnamefont{and}
  \bibinfo{author}{\bibfnamefont{A.}~\bibnamefont{Derevianko}},
  \bibinfo{journal}{Phys. Rev. A} \textbf{\bibinfo{volume}{74}},
  \bibinfo{pages}{020502(R)} (\bibinfo{year}{2006}).

\bibitem[{\citenamefont{{Pereira Dos Santos}
  et~al.}(2002)\citenamefont{{Pereira Dos Santos}, Marion, Bize, Sortais,
  Clairon, and Salomon}}]{PerMarBiz02}
\bibinfo{author}{\bibfnamefont{F.}~\bibnamefont{{Pereira Dos Santos}}},
  \bibinfo{author}{\bibfnamefont{H.}~\bibnamefont{Marion}},
  \bibinfo{author}{\bibfnamefont{S.}~\bibnamefont{Bize}},
  \bibinfo{author}{\bibfnamefont{Y.}~\bibnamefont{Sortais}},
  \bibinfo{author}{\bibfnamefont{A.}~\bibnamefont{Clairon}}, \bibnamefont{and}
  \bibinfo{author}{\bibfnamefont{C.}~\bibnamefont{Salomon}},
  \bibinfo{journal}{Phys. Rev. Lett.} \textbf{\bibinfo{volume}{89}},
  \bibinfo{pages}{233004} (\bibinfo{year}{2002}).

\bibitem[{\citenamefont{Szymaniec et~al.}(2007)\citenamefont{Szymaniec,
  Chalupczak, Tiesinga, Williams, Weyers, and Wynands}}]{SzyChaTie07}
\bibinfo{author}{\bibfnamefont{K.}~\bibnamefont{Szymaniec}},
  \bibinfo{author}{\bibfnamefont{W.}~\bibnamefont{Chalupczak}},
  \bibinfo{author}{\bibfnamefont{E.}~\bibnamefont{Tiesinga}},
  \bibinfo{author}{\bibfnamefont{C.~J.} \bibnamefont{Williams}},
  \bibinfo{author}{\bibfnamefont{S.}~\bibnamefont{Weyers}}, \bibnamefont{and}
  \bibinfo{author}{\bibfnamefont{R.}~\bibnamefont{Wynands}},
  \bibinfo{journal}{Phys. Rev. Lett.} \textbf{\bibinfo{volume}{98}},
  \bibinfo{pages}{153002} (\bibinfo{year}{2007}).

\bibitem[{\citenamefont{Rehse et~al.}(2004)\citenamefont{Rehse, Bockel, and
  Lee}}]{RehBocLee04}
\bibinfo{author}{\bibfnamefont{S.~J.} \bibnamefont{Rehse}},
  \bibinfo{author}{\bibfnamefont{K.~M.} \bibnamefont{Bockel}},
  \bibnamefont{and} \bibinfo{author}{\bibfnamefont{S.~A.} \bibnamefont{Lee}},
  \bibinfo{journal}{Phys. Rev. A} \textbf{\bibinfo{volume}{69}},
  \bibinfo{pages}{063404} (\bibinfo{year}{2004}).

\bibitem[{\citenamefont{Camposeo et~al.}(2006)\citenamefont{Camposeo, Marago,
  Fazio, Kloter, Meschede, Rasbach, Weber, and Arimondo}}]{CamMarFaz06}
\bibinfo{author}{\bibfnamefont{A.}~\bibnamefont{Camposeo}},
  \bibinfo{author}{\bibfnamefont{O.}~\bibnamefont{Marago}},
  \bibinfo{author}{\bibfnamefont{B.}~\bibnamefont{Fazio}},
  \bibinfo{author}{\bibfnamefont{B.}~\bibnamefont{Kloter}},
  \bibinfo{author}{\bibfnamefont{D.}~\bibnamefont{Meschede}},
  \bibinfo{author}{\bibfnamefont{U.}~\bibnamefont{Rasbach}},
  \bibinfo{author}{\bibfnamefont{C.}~\bibnamefont{Weber}}, \bibnamefont{and}
  \bibinfo{author}{\bibfnamefont{E.}~\bibnamefont{Arimondo}},
  \bibinfo{journal}{Appl. Phys. B} \textbf{\bibinfo{volume}{85}},
  \bibinfo{pages}{487} (\bibinfo{year}{2006}).

\bibitem[{\citenamefont{Lurio and Prodell}(1956)}]{LurPro56}
\bibinfo{author}{\bibfnamefont{A.}~\bibnamefont{Lurio}} \bibnamefont{and}
  \bibinfo{author}{\bibfnamefont{A.~G.} \bibnamefont{Prodell}},
  \bibinfo{journal}{Phys. Rev.} \textbf{\bibinfo{volume}{101}},
  \bibinfo{pages}{79} (\bibinfo{year}{1956}).

\end{thebibliography}

\end{document}